# SYMMETRY AND TRANSFORMATION OF WAVES IN OPTICS AND ACOUSTICS OF CRYSTALS

A.G.Khatkevich, L.A. Khatkevich

Belarusian State University

## ABSTRACT

*It is show that in group representation by **non-traditionally** determining by the Maxwell equations, instead of wave, linear differential operator of momentous type from the common point of view the transformation of electromagnetic and ultrasonic radiation as well as the formation of caustics generation of solitons in crystals is represented. It is established that forming operator structural constants determine bias current with the connected charge and group velocity and also optical and acoustic axes of a crystal, which characterize its wave properties, moreover crystals are classified on common electromagnetic base. It is discovered that at change of crystal symmetry and representation of different wave process the problems also take place, which are similar to others spheres of physics and are constructed on the same aximatical base.*

**1. Introduction.** The symmetry of electromagnetic field in a crystal is mathematically correct taken into account by the developed in the theory of numbers, algebraic equation and geometry theory of groups, which has been used initially in crystallography [1]. Then this theory has effectively been used at creating special and general relativity theory [2], quantum theory [3] and in the middle of the previous century the relativistic quantum theory of interacting fields [4]. At present time on nonlinear phenomena on this basis, particularly, diffraction and dispersion of waves, are intensively studied [5,6] and also different representations of the theory of groups as well as their application are developed [7-9]. The group vector representation of wave processes in crystals based on phenomenological electrodynamics is developed in [10 – 14]. Here based on this simple representation of continuous groups, the interaction and transformation of radiation is studied in optics, acoustics and acousto-optic of crystals as well as in physics of solid.

**2. Group representation of the wave processes in crystals.** The change of electromagnetic field is described by the Maxwell equations:
$$\nabla^\times \mathbf{E} + \partial \mathbf{B} = 0, \ \nabla \mathbf{B} = 0, \ \nabla^\times \mathbf{H} - \partial \mathbf{D} = \mathbf{I}, \ \nabla \mathbf{D} = \rho, \tag{1}$$
where $\mathbf{E}, \mathbf{D}, \mathbf{H}$ and $\mathbf{B}$ are the strength and induction vectors of electric and magnetic field, $\rho$ and $\mathbf{I}$ are the densities of free charge and conduction current, $\partial = \partial/\partial t$ is the time derivative, $\nabla = \partial/\partial \mathbf{r}$ is the vector of spatial derivatives and $\nabla^\times$ is the equivalent dual to it tensor $(\nabla^\times)_{ik} = \delta_{ijk}\nabla_j$, and $\delta_{ijk}$ is the Levi-Civita tensor. At retaining the charge $\partial\rho + \nabla\mathbf{I} = 0$, the system (1) is combined and at $\rho = \mathbf{I} = 0$ magnetoelectric symmetry takes place.

Eqs. (1) are supplemented with material equations (coupling equations) $\mathbf{D} = \varepsilon\mathbf{E}$ and $\mathbf{B} = \mu\mathbf{H}$ characterizing properties of crystals, where $\varepsilon$ and $\mu$ are the real symmetric and diagonolized tensors of relative dielectric and magnetic permeability of dia-, para- and ferromagnetic and, particularly, dielectrics at $\mu_{ik} = \mu_0\delta_{ik}$, where $\delta_{ik}$ is the Kronecker tensor, and also $\delta_{ijk}\delta_{lmk} = \delta_{il}\delta_{jm} - \delta_{im}\delta_{jl}$. In group representation by the introduction of tensors of impermeability roots $\varepsilon^{-1/2} = A$, $\mu^{-1/2} = B$ the vector equations (1) symmetrize and directly determine the operator $\partial \pm J$, where $J = A\nabla^{\times}B$ is the **linear differential operator of momentous type** [10,11]. Here the radiation propagation in a crystal is given in the form of closed transformation (group) of rotation the electric field into temporal change of magnetic one and vice versa. Moreover, the following generalization is possible.

Homogeneous Maxwell equations satisfy identically the ratios $\mathbf{B} = \nabla^{\times}\mathbf{\Phi}$ and $\mathbf{E} = -(\partial\mathbf{\Phi} + \nabla\Phi_4)$, which introduce 4-vector of potential $\Phi = (\mathbf{\Phi}, \Phi_4)$ for presentation of electromagnetic radiation. This 4-vector is determined by its calibration with the accuracy up to scalar $\Phi$ satisfying the wave equation. At that, the properties of crystals and anisotropic media are fuller taken into account by the following material equations:

$$\mathbf{H} = \mu^{-1}(\mathbf{B} - \kappa\mathbf{E}) = \mu^{-1}[\nabla^{\times}\mathbf{\Phi} + \kappa(\partial\mathbf{\Phi} + \nabla\Phi_4)] = \mu^{-1}(\mathbf{B} - \mathbf{M}),$$
$$\mathbf{D} = \varepsilon\mathbf{E} - \lambda\mathbf{H} = \varepsilon'(\partial\mathbf{\Phi} + \nabla\Phi_4) + \lambda\mu^{-1}\nabla^{\times}\mathbf{\Phi} = \varepsilon\mathbf{E} + \mathbf{P}, \qquad (2)$$

where $\varepsilon' = \varepsilon - \lambda\mu^{-1}\kappa$ and $\kappa$ and $\lambda$ are the tensors of magnetoelectric and electromagnetic susceptibility, which determines the vectors of crystal magnetization $\mathbf{M} = \mathbf{B} - \mu\mathbf{H} = \kappa\mathbf{E}$ and its electropolarization $\mathbf{P} = \mathbf{D} - \varepsilon\mathbf{E} = \lambda(\mathbf{H} - \mu^{-1}\kappa\mathbf{E})$ depending quadratically on susceptibilities.

Inhomogeneous Maxwell equations, vectors of which are determined by Eqs. (2), are transformed into wave equations for potentials. In group representation these wave equations in crystal optics, and in the result also in crystal-acoustics, symmetrize and factorize [11,12] and at coulomb calibration of potential, when $\Phi_4 = 0$, reduce to existing (1) vector equations: $\Lambda\Psi \equiv (\partial - L)\Psi = \mathbf{I}'$, $\Lambda^+\Psi^* \equiv (\partial + L^+)\Psi^* = \Psi$, wherein $^+$ means they selfcoupling or Hermitian character, and there are introduced complex vectors of potential $\Psi = A^{-1}\mathbf{\Phi}$, current density $\mathbf{I}' = A^{-1}\mathbf{I}$ and instead of J more general linear operator of momentum type L:

$$\Lambda = (\delta_{ik} + A_{il}\kappa_{lr}B_{rk})\partial - A_{il}\delta_{ljr}B_{rk}\nabla_j = \partial - J + K = \partial - L, \qquad (3)$$

here the wave operator is $\Lambda\Lambda^+ = \Lambda^+\Lambda = \partial^2 - LL^+$.

When using the solution (1) in a form of harmonic waves of potential $\Psi = \mathit{\Psi}\exp i(\omega t + \mathbf{kr})$, where $\mathit{\Psi}$ is the amplitude, $\omega$ is the cyclic frequency, $\mathbf{k} = \omega\mathbf{m} = \omega n\mathbf{n} = k\mathbf{n}$ is the wave vector ($\mathbf{m}$ is the refraction vector, n is the refraction index and $\mathbf{n}$ is the wave normal), derivatives $\partial$ and $\nabla$ appear to be imaginary numbers $i\omega$ and $i\mathbf{k}$. Thereby differential equations transform into algebraic ones, the operator – in a tensor, and the investigation of radiation in crystals is aggregated to the study of representation of groups of continuous transformations based on the vector algebra (**Lie algebra**). Here related to the frequency Fourier-image of the operator $L$ appears to be a tensor:

$$L/\omega = A_{il}\delta_{ljr}B_{rk}m_j - A_{il}\kappa_{lr}B_{rk} = J - K = U, \qquad (4)$$

satisfying the unitarity eqaution $UU^+ = 1$.

Thus, in group representation the Maxwell equations (1) symmetrize and determine the operator of spatial derivatives of **crystal-optics (the field)** J with the operator of temporal derivatives of **interaction of a crystal with radiation** K, forming a linear operator in a general case of **crystal-acoustics (solid or medium)** $L = J - K$. Below it is shown how the change of the symmetry of the properties of a crystal and radiation determines the character (type, rank and order) of transformation groups and metrics with the singature of vector space representation and backwards

**3. Classic crystal optics.** In a non-gyrotropic magnetic crystal the operator of crystal-optics is $J = A\nabla^{\times}B = C\nabla = C_{ijk}\nabla_j$, where $C = A^{\times}B = -B^{\times}A$ is the tensor of the $3^{rd}$ rank of so-called **structural constants**, formed by impermeability the roots, and propagation of radiation is presented by linear unimodular (special) S unitary U and also orthogonal O parametric groups of the $1^{st}$ rank $A_1 \sim B_1 \sim C_1$: the group of transformation of plane or spinor (quaternion) Sp group $E(2) \sim SU(2) \sim SO(3) \sim Sp(1) \sim Q$ and linear group, locally isomorphous to substantial subgroup of complex spinor group $SL(2) \sim SU(1,1) \sim SO(2,1) \sim Sp(1,R)$. At that as structural constants are expressed by means of dyad **A•B of main vectors** (eigen values and vectors) of impermeability roots $A = A_{ii}\mathbf{a}_i•\mathbf{a}_i = \mathbf{A}•\mathbf{a} = \mathbf{a}•\mathbf{A}$ and $B = \mathbf{B}•\mathbf{b} = \mathbf{b}•\mathbf{B}$, **tensor orthogonal operator** $J = A\nabla^{\times}B = [\mathbf{A}•\mathbf{B}]\nabla$ appears to be dual to **4-vector operator** of energy-pulse consisting of antisymmetric vector component $[\mathbf{AB}]\nabla$ and symmetric scalar $(\mathbf{AB})\nabla$ one.

In analogous way the realizing rotation and satisfying the substantial ratio $OO^T = OO^{-1} = 1$, orthogonal tensor $O = J/\omega = [\mathbf{A}•\mathbf{B}]k/\omega$ and tensor (dyad) of velocities $V = J/k = [\mathbf{A}•\mathbf{B}]\mathbf{n}$ are expressed**.** In the result vectors of ray (group) velocities of isonormal waves are represented in the following way. By dual to the antisymmetric dyad component of vector product or **root vector** (commutator) $[\mathbf{AB}] = \mathbf{u}$ the average velocity is determined of linearly polarized isonormal waves, and by symmetric in a form of scalar product (anticommutator) $(\mathbf{AB}) = u_4$ – their difference, i.e. birefringence determined by anisotropy due to **bound charge** and its rotation, namely, spin.

Waves like cyclic rotations moves faster and (or) slower with velocities, which are presented parametrically by 4-vector $\mathbf{u} \pm u_4$ and vector-parameter $\mathbf{u}/u_4$ [7]. Wave (beam) surface appears to be the surface of ellipsoid stratified into two-cavity surface. On special points of surface $u_4 = 0$ and determined by root vector **optical axes** there are distinguished positive and negative uniaxial dielectrics and magnetic crystals with the symmetry axis higher than the $2^{nd}$ order $\mathbf{e_3}$ and biaxial ones, wherein along the orthogonal $\mathbf{e_3}$ so-called "knots lines", in a form of **split** optical axes (binormals $\mathbf{n}_0$ and biradials $\mathbf{s}_0$) the conical refraction is observed.

**4. Gyrotropy and wave or quantum optics.** Interaction of radiation with eigen gyrotropic crystal excites transversal **displacement current,** due to which imper-meability tensors ermitize, structural constants $C_{ijk}$ appear to be complex and the operator J determines the unitary 8th parametric group of the 2nd rank $A_2 \sim SU(3)$, which is represented by three-dimensional matrices and, particularly, the known λ- Gell-Mann matrices [8]. In the result in eigen (or by force) gyrotropic crystals the cyclic rotation with movement transforms into helical motion in a form of left- and right-circularly and elliptically polarized waves. Here the **additional rotation** eliminates optical axes and multiple (conical) points, separates hollows of the velocity surface and radiation in a cone of beams of external conical refraction can make self-focusing.

From the microscopic point of view [11,12] in a non-gyrotropic crystal the radiation is linearly polarized and "fermionized" due to the charge and spin and is represented by orthogonal operator and asymmetric velocity tensor. In eigen gyrotropic crystals the radiation is "bosonized" due to additional rotation and the operator appears to be **unitary** and tensor and velocity dyad *V* to be **hermitian one**.

According to the quantum theory main vectors become the operators of production **u\*** and destruction **u**, the components of which satisfy commutation relation $[u_i u_k] = \delta_{ijk} u_j$. Here only one component of the group velocity is determined, for example, $u_3$ and velocity square is expressed by the equation $\mathbf{u}^2 = \pm [\mathbf{uu^*}] + u_3(u_3 \pm 1)$. The elimination of the optical axes is accompanied with quantization of radiation and transformation of 4-vectors of velocities $\mathbf{u} \pm u_4$ into quaterion $u_4 \pm i\mathbf{u}$ and bispinor $u_4 \pm \mathbf{u}\sigma$, where $\sigma_j$ is the Pauli matrices, satisfying the ratio $\sigma_i \sigma_k = \delta_{ik} + i\delta_{ijk}\sigma_j$. Invariants (squares) of 4-vector (bispinor) and quaterion: $\mathbf{u}^2 - u_4^2 = W$ и $\mathbf{u}^2 + u_4^2 = H$ appear to be different and are connected with Lagrangians and Hamiltonians formed by the square of the angular momentum. It should be noted that by the equation $\nabla^{\times 2} = \nabla \bullet \nabla - \nabla^2$ ($\delta_{ijk}\nabla_j \delta_{klm}\nabla_l = \nabla_i \nabla_m - \nabla^2 \delta_{im}$) it is introduced the dyadic (tensor) product and **metric tensor** $C^j_{ik} C^k_{jl} = g_{il}$, here $C^l_{ik} g_{lj} = C^l_{ik}\delta_{lj} = C_{ijk} = -C_{kji}$.

**5. Optical activity and nonreciprocity.** In optically active crystals (with spatial dispersion or natural gyrotropy) in orthogonal and at all unitary operator J the complex vector of derivatives $\nabla' = \nabla + i\kappa\partial$ appears, where κ is the **gyration vector**, which is determined by imaginary component of the Hermitian tensor $\kappa = \kappa' + i\kappa^{\times}$. In these crystals the propagation of radiation is being active in 4-dimensional Minkowski spacetime 6-parametric Lorentz group $D_2 \sim SO(3,1)$, locally isomorphic noncompact subgroup $SU(1,1)$ of complex group $SL(2C)$. The optically active crystal magnetizes and polarizes electrically; the operator and tensor completize also, i.e. cyclic rotation **is supplemented with spatial-temporal vibration** with obtaining phase difference by waves.

In natural as well as in eigen gyrotropic crystals $u_4/u \cong 10^{-2}$ linear operator L = J − K = J′ − K′, where Aκ′B∂ = K′ = 0, comes to operator $J' = A(\nabla + i\kappa\partial)^{\times} B \equiv A\nabla'^{\times}B$, wherein in a common case the transverse bias current unites with current determining by the gyration vector κ. Then in a general case of gyrotropic crystals the parameters of contradirectional waves appear to be different: in magnetoelectricians at $\kappa' \cong 10^{-4}$ there is the phenomenon of **optical nonreciprocity**, taking into account what the radiation is represented by quaternions or bispinors "quantizes secondary".

Due to gyration vector **κ** by linear operator in a crystal at the fulfillment of the condition of synchronism $\mathbf{k}_\pm' \pm \mathbf{k}_\pm'' = \mathbf{k}$ nonlinear transformation of frequency $\omega' \pm \omega'' = \omega$ [10,13] realizes in ferromagnetics, antiferromagnetics and ferrites together with complexification and $\kappa \approx \kappa' \geq \kappa^\times$ "isotropic spin" is taken into account with the exchange interaction of quasilongitudinal spin waves and in common case in a solid at $\lambda = \lambda^+ \geq \kappa$ – "hypercharge" with the mass and ultrasound and thermal waves [13,14].

**6. Waves in magnetic-electricians and ferromaentics.** In ferromagnetic electromagnetic perceptibility differs by the value of tensor κ, and in a solid and, particularly, in ferroelectric is replaced by the tensor $\lambda \geq \kappa$, which changes the symmetry with the creation of the possibility of introduction of so-called **Christoffel symbols** $\Gamma_{ijk} = \Gamma_{kji} = \Gamma_{i\,k}^{l} g_{lj} = \Gamma_{i\,k}^{l} \delta_{lj}$. Then by the operator L at $\kappa \neq 0$ and $K' = 0$ affine transformation is represented in a form of rotation with the **shift deformation,** and at $\lambda' \geq \kappa' \neq 0$ and $K' \neq 0$ with the shift deformation and stress-strain. In the result in ferromagnetics nonlinear polarized quasitransverse and quasilongitudinal spin waves (magons) and in a case of a solid – ultrasound waves (photons) are excited and the interaction and transformation takes place of electromagnetic, spin and ultrasound waves.

In a common case dyad **A•B** is added with the third independent (noncoplanar) representing $\lambda' \geq \kappa'$ be the resultant vector **C**, and structural constants $C_{ijk}$ (and $\Gamma_{ijk}$) are formed by three vectors, which are connected with base vectors of the crystalline grating and crystal physical basis. By the dyad of noncollinear vectors[**AB**]•**C** vector product with Jacobia associativity [[**AB**]**C**] + [[**BC**]**A**] + [[**CA**]**B**] = 0, as well as scalar (mixed) product with gramian $([\mathbf{AB}]\mathbf{C})^2 \neq 0$ determined, which after vector normalization is expressed in the following form: $([\mathbf{ab}]\mathbf{c})^2 = [\mathbf{ab}][\mathbf{bc}][\mathbf{ca}] = \mathbf{a}^2\mathbf{b}^2\mathbf{c}^2 - \mathbf{a}^2(\mathbf{bc}) - \mathbf{b}^2(\mathbf{ca}) - \mathbf{c}^2(\mathbf{ab}) + 2(\mathbf{ab})(\mathbf{bc})(\mathbf{ca})$, where $\mathbf{a}^2 = \mathbf{b}^2 = \mathbf{c}^2 = 1$.

In vector representation the change of the symmetry of crystal properties is determined by the structural constants, which form the operator L, tensor $U$ and corresponding algebra, for example, of quaternions:

$$\mathbf{u}'' = \mathbf{u} \cdot \mathbf{u}' = (\mathbf{u} + i\mathbf{u}_4) \cdot (\mathbf{u}' - i\mathbf{u}'_4) = [\mathbf{u}\mathbf{u}'] + (\mathbf{u}\mathbf{u}') + i(\mathbf{u}_4\mathbf{u}' - \mathbf{u}'_4\mathbf{u}') + \mathbf{u}_4\mathbf{u}'_4. \quad (5)$$

In this expression to double numbers corresponding to the components of $3^{rd}$ tensor imaginary and dual numbers (Clifford-Lipshitz hypercomplex numbers) are added and $4^{th}$ tensors corresponding to the operators J′ and L and Lorentz and Puankare transformation [10-14] are formed. Invariants of these 4-tensors appear to be coefficients of the $4^{th}$-order equation, which is solved in radicals using the Decart- Euler or Ferrari method.

The change of vector field in a solid in more common case of 5-dimensional space is studied using the Thomas–Fermi or Hartree–Fock method and transformation and agreement of electromagnetic and ultrasound fields are being half-simple orthogonal and symplectic (simple) groups of the $2^{nd}$ rank of $10^{th}$ order of $B_2 \sim C_2$: SO(5) ~ Sp(2) or SO(3,2) ~ Sp(2,R) and SO(4,1) ~ Sp(1,1), respectively. These groups are characterized by the track and determinant and here the real group $D_2 \sim A_1 \cdot A_1$ (but not $D_2 \sim SO(3,1)$!) is not a simple one.

**7. Relativistic quantum theory.** When taking into account the phenomenon of nonreciprocity the Maxwell equations for complex potential appear to be similar to relativistic Dirac equations [10-12]. Here 4-vectors (quaternions and bispinors) transform into biquaternions (octaves) and appear to be 4 tensors represented by α and β or γ-matrices satisfying the ration $\gamma_\alpha \gamma_\beta + \gamma_\beta \gamma_\alpha = 2\delta_{\alpha\beta}$, $\alpha, \beta = 1,...4$, and forming **Dicar algebra** having 32 element and 17 classes.

In group vector representation the radiation transformation is determined (similar to a multiplication table) by material equations (2), wherein 36 substantial parameters determine 27 components of the tensor of $3^{rd}$ rank $C_{ijk}$ formed by hermitian tensors, which in classic crystal physics come to $18^{th}$ parametric complex tensor $U$ similar to the Petrov tensor in relativistic cosmology [9]. At symmetrization groups of the $3^{rd}$ rank of $15^{th}$ order $D_3 \sim A_3$ and $21^{st}$ order $B_3$ and $C_3$ are naturally introduced, which are represented by the $18 \pm 3$ parameters, i.e. 15 in linear optics and 21 in acoustics of crystals [13,14].

It should be noted that in vector presentation of $C_{ijk}$ and $\Gamma_{ijk}$ dimensions of the introduced spaces and number of independent elements are connected by the ratio $3^2 + 4^2 = 5^2$. Here the crystals on the presence of rotation axes $C_n$ and with inversion $S_n$, $n \geq 1, 2, 3, 4$ and 6, and plains of symmetry $D_n$, $n \geq 1, 2, 3$, are divided into 32 crystallographic classes [1] and 7 systems (subgroups of complex rotation group [7]) with 14 different Bravais cells, here there are 11 types of naturally gyrotropic crystals [15], as well as types of anisotropic homogeneous spaces [9].

The equation (5) after normalization comes to the ratio of orthogonality and phase-beam symmetry $(u/|u|) \cdot (u'/|u'|) = UU^+ = VN = E$, where $E$ is the product of tensor $\delta_{ik}$ and δ-function, here quasi-inverse to velocity tensor $V$ tensor of indices $N$ appears to be the Green function G – fundamental solution of corresponding (1) system of equations for the potential [13,14]. This equation combines vector product [**um**] − compatibility and integrability condition − with Jacobi associativity, zero curvature of space and spin and **isospin of quasiparticles,** and scalar (**um**) – with metrics and signature of space and **hypercharge with mass** of particles and antiparticles.

It should be noted that at elimination of binormals there appears the energy quant $\hbar = h/2\pi = 1{,}054 \cdot 10^{-27}$ *erg•s*, where h is the Planck's constant, and 4-vector connected charge and bias current in a form of ratio of Faraday and Avogadro constants $\varepsilon_0 = F/A = (137{,}04)^{-1/2}\kappa$ and spin $\mu_0 = 0{,}5008\mu_B$, where $\mu_B = \hbar\varepsilon_0/2m_0c$ is the Bohr magneton determined by the charge $\varepsilon_0 = 1{,}602 \cdot 10^{-19}\kappa$ and electron mass $m_0 = 9{,}110 \cdot 10^{-28}$ *e* at velocity c = $2{,}998 \cdot 10^{10}$ *cm/s* and $\mu_0\varepsilon_0 c^2 = 1$.

**8. Crystal acoustics.** Eigen values of tensor $L$ are determined by the characteristic equation:

$$\lambda^3 - 3a\lambda^2 + b\lambda - c = 0, \text{ or } \lambda'^3 + 3p\lambda' + q = 0, \tag{6}$$

where real coefficient $a = \Sigma\lambda_i$, $b = \Sigma\delta_{ijk}\lambda_j\lambda_k$ and $c = \lambda_i\lambda_j\lambda_k$ are the invariants of tensor, and $\lambda' = \lambda - a$, $p = b - a^2$ and $q = c - ab + a^3/2$. Equation (6) at positive or negative and equal to discriminant $p^3 + q^2$ has real and two complex conjugated roots or three real roots, at which two, and at $q = p = 0$ even three roots coincide. It is essential that at complex coefficients Eq. (6) is soluble in radicals, if only the root $\lambda_3$ is a rational one. The equation of the 4$^{th}$ stage comes to cubic (its resolvent) with the same discriminant and is solved in radicals using the above-pointed method.

In crystal acoustics ($[\mathbf{AB}]\mathbf{C}) \neq 0$, and usually by Eqs (2) in a form of matrix with 21 symmetric modulus of elasticity $c_{ijkl} = c_{аб}$, $a, б = 1,2,..6$, three orthogonal linear polarized isonormal elastic ultrasound waves are determined. In crystals these waves have also two equal velocities like in isotropic medium, but in special directions of **acoustics axes** [16]. At coincidence of $\mathbf{n}_0$ with the direction of extreme velocities and polarization vectors with unit vectors $\mathbf{e}_i$ in triclinic crystals only 18 independent (invariant) modules remain, which are determined by substantial constants in Eq. (2). In monoclinic crystals at coincident of unit vector, for example, $\mathbf{e}_1$ with axis of symmetry of 2$^{nd}$ order or direction orthogonal to the symmetry plane, 12 modules remain.

At wave normal in symmetry plane of monoclinic crystal of the components $L_{12} = L_{13} = 0$, the cubic equation is divided into equations of the 1$^{st}$ and 2$^{nd}$ orders, and the velocity of the transverse wave is determined by the component $L_{11} = \lambda'_1$, and two other by quadric equation $a'\lambda'^2 + b'\lambda' + c' = 0$, where $b'/a' = \lambda'_2 + \lambda'_3 = L_{22} + L_{33}$ and $c'/a' = \lambda'_2\lambda'_3 = L_{22}L_{33} - L_{23}^2$ are the invariants and at $(b'/2)^2 - a'c' = 0$, i.e. at corresponding to acoustic axes ones of symmetry $\lambda'_3 = \lambda'_2$. Here in more symmetric crystals in the symmetry plane there are formed corresponding to the order of symmetry axes "strips" of extreme directions of velocities. In rhombic crystals 9 invariant modules of 12 remain, in tetragonal and trigonal 6, and in hexagonal and cubic 5 and 3, respectively.

In a cubic crystal the 3$^{rd}$ parameter $c_{12} \neq c_{11} - 2c_{44}$ appears, which characterizes the anisotropy with the different sign like "non-euclidicity" and there are three symmetry axes of the 4$^{th}$ and four of 3$^{rd}$ order. In hexoganal crystal $c_{12} = c_{11} - 2c_{66}$ and anisotropy are characterized by quadratic invariant $(c_{33} - c_{44})^2 - (c_{23} - c_{44})^2 = (c_{33} + c_{23})(c_{33} - 2c_{44} - c_{23}) \neq 0$ and, consequently, by the sign of parameter $c_{23} \neq c_{33} - 2c_{44}$. Tetragonal and trigonal crystal has additionally 6$^{th}$ invariant module, $c_{12} \neq c_{11} - 2c_{66}$ and $c_{14} \neq 0$, respectively. In a rhombic crystal the anisotropy differs by the signs of 3 of 9 symmetric modules $c_{аб}$, $a \neq б$ [16].

In a crystal acoustics together with excitation of ultrasound waves the interaction and conserved values change. Electromagnetic energy and **pulse momentum** represented in crystal optics by Lagrangians $W = \mathbf{u}^2 - u_4^2$ and Hamiltonians $H = \mathbf{u}^2 + u_4^2$ and exceptional groups $G_2$ and $F_4$ connected with preserving **energy and charge** are replaced by mechanical kinetic and potential (elastic) energy and **angular momentum**. Here different interactions of electromagnetic and ultrasound waves with vibrations of elements in a form of tetrahedron (rhomb), cube (octahedron) and icosahedrons (dodecahedron) are represented by exceptional groups $E_6$, $E_7$ and $E_8$. These interactions are connected with invariants of tensor $L$, which are determined by length, square with the curvature of surface and value in geometry, and by energy, interacting and inertial change and mass in physics.

**9. Conical refraction, caustics and solitons.** Beams and pulses of waves are represented by the diffraction integral $\psi(\mathbf{r},t) = \int dt'\mathbf{dr}' G(\mathbf{R},T) \psi(\mathbf{r}',t')$, where $G(\mathbf{R}, T) = \exp[i(\mathbf{kR} \pm \omega T)]$ is the Green function, $\mathbf{R} = \mathbf{r} - \mathbf{r}'$, $T = t - t'$. Cyclic frequency is expressed in the form of [16-21]:

$$\omega = \omega_0 + \mathbf{uq} + \mathbf{q}w\mathbf{q}/2, \tag{7}$$

where $\mathbf{q} = \mathbf{k} - \mathbf{k}_0$, $\mathbf{k}_0$ is the wave vector of central wave, and $\mathbf{u} = d\omega/d\mathbf{k}$ and $w = d\mathbf{u}/d\mathbf{k}$ are the group velocities and curves at different $\mathbf{k}_0$ and $\omega_0$ for various hollows of velocity surface.

When using Eq. (7) the index of the Green tensor function $G = U^+$ transforms to the difference of quadratic forms in a form of Lagrangian: $W = \mathbf{q}(wt)\mathbf{q} + 2\mathbf{u}'\mathbf{q} = \mathbf{q}'(wt)\mathbf{q}' - \mathbf{u}'(wt)^{-1}\mathbf{u}'$, where $\mathbf{q}' = \mathbf{q} + \mathbf{u}'$ и $\mathbf{u}' = \mathbf{u}t \pm \mathbf{r}$. By this difference of quadratic form the entropy $S = k \ln W$ is presented, where $k = 1{,}38 \cdot 10^{-9}$ *joule/deg.* is the Boltzmann constant, which is determined by the expression $k = R/A$ by gas constant $R = 8{,}314$ *joule/deg.* $= 1{,}986$ cal/*deg.* and the Avogadro constant. From the other side $S = Q/T$, where Q is the energy losses in a form of heat and $T = 273$ is the absolute temperature connected with the constant of thin structure $\alpha$ by the equation $(2 \cdot 137) - 1 = \alpha/2 - 1$.

The Eq. (7) approximates the hollows of the surface of velocities by the 2-nd order surfaces, the peculiarities of which are determined by determinant $|w|$ and discriminant $\mathbf{q}w\mathbf{q} + (\omega_0 \pm \omega)|w|$. In eigen basis $\mathbf{e}_i$ tensor diagnolizes, $|w| = w_{III} = w_1 w_2 w_3$ and are represented by the ellipsoid, which connects basis $\mathbf{e}_i$ with the symmetry of crystal and determines the properties of hollows of its stratified velocity surface by optical and acoustic axes. These hollows of velocity surface is characterized by full (Gaussian) curvature $w_{II} = w_1 w_2$ and average curvature $w_I = (w_1 + w_2)/2$. Depending on $w_{II} > 0$ or $w_{II} = 0$ and $w_{II} < 0$ points of hollows are elliptical (double) or special parabolic (dual) ones with multiple (conical) one at $w_1 = w_2 = 0$ and hyperboic (saddle), and directions are ordinary or special ones.

In a case of ordinary direction, since $G = \Pi^3_{j=1} G_{j\pm 1}$ and the integration is fulfilled using the ratio $(\pi/2)^{-1/2} \int_{-\infty}^{\infty} dq'_j \exp \pm i q'^2_j = \pi^{1/2}(1 \pm i)/2$, it appears that the divergence of wave beams changes in time inversely proportional to the curvature of the surface of velocities. At that, beams and pulses are represented in a canonic form in semigeodesic coordinates by different special, including orthogonal, functions and polynominals. In a case of special directions of optical and acoustic axes (binormals $\mathbf{n}_0$ and biradials $\mathbf{s}_0$) integrals $G_{j\pm 1}$ appears to be divergent and are studied separately.

In a case of binormals $\mathbf{n}_0$ and parabolic points of surface velocities are to be limited by first terms of $p^{th}$ (higher than $2^{nd}$) order and integrals are to be studied. Then the simplest seven catastrophes are realized with the formation of caustics, which are denoted as $A_{p-1}$, $p = 3, 4, 5, 6$ and $D_{p+1}$, $p = \pm 3, 4$, at which the wave amplitudes (quaternions or bispinors) have different dependence on $\mathbf{u}' = \mathbf{u}t \pm \mathbf{r}$ [17, 18]. Thus, the dependence is described by G-function (and factorials), the values $G_{j\pm 1/2} =$ $\pi^{1/2}\exp\pm i(\pi/2) = \pi^{1/2}(1 \pm i\pi/2)$ of which are connected with the same in ordinary directions and determine the connection of catastrophes with nonlinearity and transcendental numbers $\pi$ and e.

In a case of biradials $s_0$ and hyperbolic (imaginary) points with cone in a crystal instantones anf solitons [19,20] are formed, which due to disperse are grouped into solitons-quasiparticles with conservation of energy, charge and mass or are absorbed. These solitons are described by the solutions of the nonlinear sine-Gordon equation. Using wave equations other nonlinear equations of the soliton theory are also obtained, wherein the dispersion is compensated by nonlinearity. The solution of these equations using the methods of the Riemann–Hilber's problems and contrary dispersion task are expressed in a form of elliptic and automorphous functions [20,21].

**10. Conclusion.** Thus, in group representation by introduction of **permeability roots** the Maxwell equations symmetrize and determine instead of wave, linear differential operator of momentous type of optics and acoustics of crystals in a form of convolution of 4-vector derivatives with the tensor of the 3$^{rd}$ rank of structural constant. For harmonic waves this operator transforms into orthogonal or unitary tensor and its dual 4-vector quaternions or bispinors of group wave velocity. Structural constants formed mainly by tensor vectors of permeability and susceptibility roots, are bound with invariants of tensor of group velocities by energy, charge and mass of crystal elements and determine characterizing them wave properties either optical or acoustic axes. The developed on this basis interpretation of wave processes gives new crystal-physical interpretation of continuous transformations and opens novel directions of investigation and possibility for practical application of crystals.

## Summary

It is established that in group representation of wave process the Maxwell equations are symmetryzed and directly reduced to simple general relations, determining parameters of radiation in phenomenological as well as in quantum relativistic optics and acoustics of crystals.